\documentstyle[12pt,psfig]{article}

\begin{document} 

\title{Reanalysis of antiproton production in proton-nucleus and
nucleus-nucleus reactions at subthreshold energies\thanks{Supported by 
Forschungszentrum J\"ulich, BMBF and the Russian Foundation for
Fundamental Research}}
\author{ A. Sibirtsev$^1$, W. Cassing$^1$, G.I. Lykasov$^2$ and 
M.V. Rzjanin$^2$ \\
$^1$Institut f\"ur Theoretische Physik, Universit\"at Giessen \\
D-35392 Giessen, Germany\\
$^2$Joint Institute for Nuclear Research \\
141980, Dubna, Moscow Region, Russia} 

\maketitle

\date{ }

\begin{abstract}
We reanalyse the production of $p\bar{p}$ pairs in proton-nucleus 
and nucleus-nucleus
collisions employing novel elementary cross sections for baryon-baryon and
pion-baryon production channels based on a boson-exchange model. 
In contrast
to previous transport studies performed in the literature 
the secondary pion induced
channel is found to be most important in both p+A and A+A collisions at
subthreshold energies. A detailed comparison with 
the experimental data available
indicates that sizeable attractive $\bar{p}$ potentials in the order of
- 100 to - 150 MeV at normal nuclear matter density are needed 
to reproduce
the size and shape of the experimental spectra.
\end{abstract}

\newpage

\section{Introduction}
The production of particles at energies below the free nucleon-nucleon
threshold ('sub\-threshold production') is one of
the most promising observable for hadron selfenergies
at high nuclear densities since the particles are
produced predominantly during the compressed stage 
\cite{Bertsch88,Cassing90,Cassing90b,Aich}.
Antiproton production at energies of a few AGeV is the most 
extreme subthreshold production process and has been 
observed in proton-nucleus collisions already a few decades ago
\cite{chamberlain,elioff,dorfan}. 
Experiments at the JINR \cite{JINR} and at the BEVALAC
\cite{BEVALAC1,BEVALAC2} have provided, furthermore, 
first measurements of
subthreshold antiproton production in nucleus-nucleus collisions followed
by measurements at KEK \cite{KEK} and GSI \cite{GSI,Schroter2} with new 
detector setups. 
Various descriptions for these data have been proposed.
Based on thermal models it has been suggested that the antiproton yield
contains large contributions from $\Delta N \rightarrow \bar{p} +X$,
$\Delta \Delta \rightarrow \bar{p} + X$  and 
$\rho \rho \rightarrow \bar{p}N$
production mechanisms \cite{Koch89,ko1,ko2}.
Other models have attempted to explain these data in terms of 
multiparticle
interactions \cite{Danielewicz90}. 

Nowadays, the relative strength of the
various production channels as well as possible in-medium effects 
are most effectively
controlled by means of transport approaches 
\cite{Ko93,Batko91,Faessler,Cass92,Teis1,Teis2}.
First results of a fully relativistic transport calculation for antiproton
production including $\bar{p}$ annihilation as well as the change of the
quasi-particle properties in the medium have been reported in 
\cite{Cass92}.
There it was found that according to the reduced nucleon mass in the 
medium the threshold for $\bar{p}$-production is shifted to lower energy
and the antiproton cross section prior to annihilation becomes enhanced 
substantially as compared to a
relativistic cascade calculation where no in-medium effects are 
incorporated.
A variety of transport studies have been performed since then - 
most of them
using the parametrization of the elementary production cross section as
proposed by Batko et al. \cite{Batko91} - and confirmed the 
necessity of attractive $\bar{p}$
potentials, however, with some debate about its actual magnitude 
\cite{Ko93,Faessler,Teis2}.

More recently, the question about the elementary $\bar{p}$ production
amplitude has been addressed by Lykasov et al. in an effective boson
exchange model \cite{LRC}. Here the production cross section close to
threshold follows accurately the 4-body phase-space constraints, but is
much lower than the commonly adopted parametrization from \cite{Batko91}.
Such a reduced production cross section in baryon-baryon collisions 
would demand for even stronger
attractive $\bar{p}$ potentials if further reactions channels might be
neglected. However, as has been found in studies of $K^+$ and $K^-$
production \cite{Cass97,Brat97} the secondary pion induced channel 
might dominate very well
such that also the $\pi N \rightarrow N p \bar{p}$ production channel
has to be considered. 

The paper is outlined as follows: In Section 2 we briefly recall the main
ingredients of the relativistic transport approach used for $\bar{p}$
production before \cite{Teis2}. Section 3 is devoted 
to the calculation of the reaction
$\pi N \rightarrow N p \bar{p}$ in a boson exchange model 
and a presentation of the elementary exclusive and inclusive
cross sections. Section 4 contains a detailed comparison of the calculated
spectra with experimental data for p+A and A+A collisions with the aim
to obtain some closer constraints on the in-medium $\bar{p}$ potential, 
while a summary on the $\bar{p}$ selfenergies concludes the paper in
Section 5.

\section{Transport approach for antiproton production and propagation}

The phase-space distribution function for the antiprotons 
$f_{\bar{p}}(x,p_{\bar{p}})$
is assumed to follow an equation of motion equivalent to the 
transport equation
for baryons \cite{KLW1},
\begin{equation}
\left[ \frac{\Pi^{\mu}_{\bar{p}}}{\Pi^{0}}_{\bar{p}}
 \partial^{x}_{\mu} + \left( \Pi^{\nu}_{\bar{p}} F_{\mu\nu}^{\bar{p}}
+ m^{\star}_{\bar{p}} \left( \partial_{\mu}^{x} m^{\star}_{\bar{p}} 
\right)
\right) \partial^{\mu}_{\Pi_{\bar{p}}} \right] f_{\bar{p}}(x,\Pi_{\bar{p}}
) = I_{coll}^{\bar{p}}(x,\Pi_{\bar{p}})
\label{vlat}
\end{equation}
with
\begin{equation}
F^{\mu\nu}_{\bar{p}}  =  \partial^{\mu} U_{v}^{\bar{p} \,\nu}
-  \partial^{\nu} U_{v}^{\bar{p} \,\mu}
		  \label{fmunu}
\end{equation}
and the mass-shell constraint
\begin{equation}
\left( \Pi^{2}_{\bar{p}} - m^{\star\, 2}_{\bar{p}} \right)
 f_{\bar{p}} \left(x,\Pi_{\bar{p}} \right)
 =   0. \label{atmass3}
\end{equation}
Since the momentum dependence of the scalar and vector selfenergy 
of the antiproton are unknown,
we assume these quantities to be independent on momentum in Eq.~(1),
however, to be of different strength as compared to the nucleons, 
\begin{eqnarray}
U_{s}^{\bar{p}}(x) & = & - g_{s}^{\bar{p}} \  \sigma_H(x) \nonumber \\
U_{\mu}^{\bar{p}}(x) & = & g_{v}^{\bar{p}} \  \omega_H(x) \label{appot} ,
\end{eqnarray}
where $\sigma$ and $\omega$ are the scalar and vector meson fields in the
nuclear Lagrangian (cf. Ref. \cite{Teis2}).
Thus the effective mass and the effective momentum of  antiprotons are
given by
\begin{eqnarray}
m^{\star}_{\bar{p}} & =  m + U_{s}^{\bar{p}}(x) & =
m - g_{s}^{\bar{p}}\sigma_H(x) \label{effapm}  \\
\Pi_{\mu}^{\bar{p}} & =  p_{\mu}^{\bar{p}} - U_{\mu}^{\bar{p}}(x) & =
p_{\mu}^{\bar{p}} - g_{v}^{\bar{p}} \omega^H(x). \label{effquap}
\end{eqnarray}
According to the arguments given before the coupling
constants $g_{s}^{\bar{p}}$ and $g_{v}^{\bar{p}}$ are treated as free
parameters that should be fixed in comparison to data.

\subsection{The collision integral}

The collision term $I_{coll}^{\bar{p}}$ (r.h.s. of eq. (\ref{vlat}))
includes i) a term $I_{elast}^{\bar{p}}$, describing the elastic
antiproton-baryon scattering, ii) a term $I_{prod}^{\bar{p}}$, 
describing the
antiproton production by baryon-baryon and meson-baryon collisions, 
and iii) a term $I_{abs}^{\bar{p}}$
responsible for in-medium $\bar{p}$-absorption on baryons. 
The production of
$p\bar{p}$ pairs by meson-meson collisions can be neglected in the energy
regime of interest here since the invariant energies available 
are too small.
$I_{elast}^{\bar{p}}$ describes elastic baryon-antiproton scattering
as well as elastic antiproton-antiproton scattering. 
While this part of the
collision integral $I_{coll}^{\bar{p}}$ can be formulated similarly to
the collision term describing bayon-baryon scattering 
the other terms represent extensions to the conventional 
collision integral.

The basis for the description of the $\bar{p}$ production by baryon-baryon
collisions is the process
\begin{equation}
 B+B\rightarrow \bar{p}+p+N+N \equiv 1 + 2 \rightarrow
\bar{p} + 3+ 4+ 5, \label{prore1}
\end{equation}
for which the corresponding covariant collision integral reads 
\begin{eqnarray}
\lefteqn{I^{\bar{p}}_{prod}(x,\Pi_{\bar{p}}) =} \nonumber \\
 &&  \frac{4}{(2 \pi)^{11}}
                     \int d^{3} \Pi_{1} \, d^{3} \Pi_{2} \,
                 d^{3} \Pi_{3} \, d^{3} \Pi_{4} \, d^{3} \Pi_{5}
  \frac{m^{\star}_{1} \, m^{\star}_{2} \, m^{\star}_{3}\,  m^{\star}_{4}\,
m^{\star}_{5}\,  m^{\star}_{\bar{p}}}{\Pi^{0}_{1}\, \Pi^{0}_{2}\,
\Pi^{0}_{3}\,  \Pi^{0}_{4}  \,
                 \Pi^{0}_{5} \,  \Pi^{0}_{\bar{p}} }\nonumber \\
         &&  W(\Pi^{\mu}_{1},\Pi^{\mu}_{2} \mid \Pi^{\mu}_{3},
                \Pi^{\mu}_{4},\Pi^{\mu}_{5},\Pi^{\mu}_{\bar{p}} )
	\delta^{4}(p^{\mu}_{1} +p^{\mu}_{2}- p^{\mu}_{3} -p^{\mu}_{4} -
  p^{\mu}_{5} -\Pi^{\mu}_{\bar{p}}  )
\nonumber \\
         && \left\{ f(x,\Pi_{1}) f(x,\Pi_{2})(1 - f(x,\Pi_{3}))
         (1-f(x,\Pi_{4})) (1 - f(x,\Pi_{5})) \right\} ,
\label{collpr}
\end{eqnarray}
where $W$ is the transition probability for the reaction
$\Pi_1 +\Pi_2 \to \Pi_3+\Pi_4+\Pi_5+\Pi_{\bar p}$ in terms of 
momentum coordinates.
We have omitted the Pauli-blocking factor for the antiproton in 
the final state 
because the number of antiprotons created during a heavy-ion collision
in the subthreshold energy regime is negligible. For the same reason we
neglect the effects of the reaction (\ref{prore1}) on the phase-space
distribution function of the baryons. The production term from $\pi N$ 
collisions has a similar structure as Eq.~(\ref{collpr}) except for the
fact that particle 2 denotes a pion and the final state is 
of 3-body type (with 2 nucleons only) since the 4-momentum of the
impinging pion is used entirely for the production of the $p\bar{p}$ pair.

In the collision integral $I_{abs}^{\bar{p}}$ we do not treat all
possible annihilation reactions separately but sum up all channels in
the inclusive annihilation reaction
\begin{equation}
B + \bar{p}  \rightarrow  X,  \label{absrea1}
\end{equation}
where X denotes all possible final states (essentially pions) 
of the baryon-antiproton
annihilation. The corresponding energy and momentum conservation reads 
\begin{equation}
p_{B}^{\mu} + p_{\bar{p}}^{\mu} = p^{\mu}_X, \label{enmomabs}
\end{equation}
where $ p_{B}^{\mu}$ and $ p_{\bar{p}}^{\mu}$ denote the 4-momenta of the
baryon and the antiproton, respectively, and $p^{\mu}_{X}$ denotes the
sum of the 4-momenta of all particles in the final state of the
annihilation reaction. Reaction (\ref{absrea1}) then leads to the
collision term
\begin{eqnarray}
I_{abs}^{\bar{p}}(x,\Pi_{\bar{p}}) & = & - \frac{4}{(2 \pi)^{3}}
                     \int d^{3} \Pi_{1}  d^{4} \Pi_{X}
                   \frac{m^{\star}_{1}\,  m^{\star}_{\bar{p}}}{\Pi^{0}_{1}
        \,     \Pi^{0}_{\bar{p}} }\nonumber \\
	 &   &	W(\Pi^{\mu}_{1},\Pi^{\mu}_{\bar{p}} \mid \Pi^{\mu}_{X})
	\delta^{4}(p^{\mu}_{1} +p^{\mu}_{\bar{p}} - p^{\mu}_{X})
	 	f(x,\Pi_{1}) f(x,\Pi_{\bar{p}})
         \label{collabs3}
\end{eqnarray}
with $W(\Pi^{\mu}_{1},\Pi^{\mu}_{\bar{p}} \mid \Pi^{\mu}_{X})$ 
denoting the
transition probability for the reaction (\ref{absrea1}). Integrating
(\ref{collabs3}) over $d^4\Pi_X$ we obtain
\begin{equation}
I_{abs}^{\bar{p}}(x,\Pi_{\bar{p}})  =  - \frac{4}{(2 \pi)^{3}}
   \int d^{3} \Pi_{1}  \frac{m^{\star}_{1}\,  
m^{\star}_{\bar{p}}}{\Pi^{0}_{1}
	\,     \Pi^{0}_{\bar{p}} } W(\Pi^{\mu}_{1},\Pi^{\mu}_{\bar{p}} )
	 f(x,\Pi_{1}) f(x,\Pi_{\bar{p}}).
         \label{collabs32}
\end{equation}
Here the integration over $d^4\Pi_X$ implies in addition to 
the integration
over all final momentum states of a particular reaction a summation
over all possible annihilation channels. Due to this fact
$W(\Pi^{\mu}_{1},\Pi^{\mu}_{\bar{p}})$ denotes the probability of an
antiproton with effective momentum $\Pi^{\mu}_{\bar{p}}$ to 
annihilate with
a baryon with effective momentum $\Pi_{1}^{\mu}$.

\subsection{Numerical implementation}

Since the production probability for antiprotons is very small 
the average time evolution of the nucleus-nucleus is not affected 
and it is
justified to treat the $\bar{p}$-production perturbatively
\cite{Cassing90,Cassing90b}. Within this approach the $\bar{p}$ invariant
differential multiplicity is obtained by summing incoherently over all
baryon-baryon and meson-baryon collisions and integrating over all 
residual degrees of
freedom. Assuming the antiproton production to take place via reactions
of the type
\begin{eqnarray}
 B+B\rightarrow \bar{p}+p+N+N \equiv 1 + 2 \rightarrow
\bar{p} + 3+ 4+ 5 \label{prore}
\end{eqnarray}
\begin{eqnarray}
 \pi+B\rightarrow \bar{p}+p+N \equiv 1 + 2 \rightarrow
\bar{p} + 3+ 4 \label{propi}
\end{eqnarray}
(B stands for either nucleon or $\Delta$) the
invariant multiplicity as a function of the impact 
parameter can be written as
\begin{eqnarray}
E_{\bar{p}}\frac{d^{3}P(b) }{d^{3}\Pi_{\bar{p}}} & = & 
     \sum_{BB coll}
 \int d^{3} \Pi'_{3} d^{3} \Pi'_{4} d^{3} \Pi'_{5} \   \frac{1}
 {\sigma_{BB}(\sqrt{s})} \
E'_{\bar{p}} \frac{d^{12}
\sigma_{BB \rightarrow \bar{p}+X}
\left( \sqrt{s} \right) }{ d^{3} \Pi'_{3} d^{3} \Pi'_{4} d^{3} \Pi'_{5}
d^{3} \Pi'_{\bar{p}}}  \nonumber \\
&   & (1- f(x,\Pi'_{3}) \ ) \  (1- f(x,\Pi'_{4}) \ ) \
 (1- f(x,\Pi'_{5}) \ ),   \nonumber \\ 
\end{eqnarray}
$$  +   \sum_{\pi B coll}
 \int d^{3} \Pi'_{3} \ d^{3} \Pi'_{4} \   \frac{1}
 {\sigma_{\pi B}(\sqrt{s})} \
E'_{\bar{p}}  \ \frac{d^{9}
\sigma_{\pi B \rightarrow \bar{p}+X}
\left( \sqrt{s} \right) }{ d^{3} \Pi'_{3} d^{3} \Pi'_{4} 
d^{3} \Pi'_{\bar{p}}}  $$
\begin{equation} (1- f(x,\Pi'_{3}) \ ) \  (1- f(x,\Pi'_{4}) \ )
,   \label{pwahr}
\end{equation}
where the quantities $\Pi_{i} \, (i=1,..,5)$ denote the in-medium
momenta of the participating baryons. $\Pi_{\bar{p}}$ and $E_{\bar{p}}$
stand for the $\bar{p}$  effective momentum and energy while 
$s=(\Pi_{1}^{\mu} +
\Pi_{2}^{\mu})^{2}$ is the squared invariant energy available in the
corresponding baryon-baryon or pion-baryon collision. 
The first sum of the Eq.~(15) is that over all baryon-baryon
collisions while the second sum is that over all pion-baryon
collisions occuring during the dynamical evolution.
An integration over the impact
parameter then yields the Lorentz invariant differential
production cross section
\begin{equation}
E_{\bar{p}} \frac{d^{3} \sigma_{\bar{p}}}{d^{3} \Pi_{\bar{p}}}
 =  2 \pi  \int d b  \: b \ E_{\bar{p}} \frac{d^{3} P(b)}{d^{3} 
\Pi_{\bar{p}}}.
\label{diffwqpr}
\end{equation}

While in free space the threshold for the elementary 
production reaction is
4 x the nucleon restmass one has to take into account the
selfenergies of all participating particles for a production 
process in the medium.
The conservation of energy and momentum has to be guaranteed, 
i.e. in baryon-baryon
reactions
\begin{equation}
p_{1}^{\mu} + p_{2}^{\mu} = p_{3}^{\mu} + p_{4}^{\mu} + p_{5}^{\mu} +
  p_{\bar{p}}^{\mu},
\label{prodenim}
\end{equation}
which in terms of effective momenta and effective masses
(cf. eqs. (5) and (\ref{effquap})) reads
\begin{eqnarray}
\Pi^{\mu}_{1} + U_{v}^{\mu}(\mid \vec{p}_{1}\mid,x) +
\Pi^{\mu}_{2} + U_{v}^{\mu}(\mid \vec{p}_{2}\mid,x)  =
\Pi^{\mu}_{3} + U_{v}^{\mu}(\mid \vec{p}_{3}\mid,x) +
\Pi^{\mu}_{4} + \nonumber \\
U_{v}^{\mu}(\mid \vec{p}_{4}\mid,x)
+\Pi^{\mu}_{5} + U_{v}^{\mu}(\mid \vec{p}_{5}\mid,x) +
\Pi^{\mu}_{\bar{p}} + U_{v}^{\bar{p} \, \mu}(x). \label{gl61}
\end{eqnarray}
With the abbreviation
\begin{eqnarray}
\Delta^{\mu}  \equiv & U_{v}^{\mu}(\mid \vec{p}_{3}\mid,x) +
		    U_{v}^{\mu}(\mid \vec{p}_{4}\mid,x) +
		U_{v}^{\mu}(\mid \vec{p}_{5}\mid,x) +
		U_{v}^{\bar{p} \, \mu}(x) \nonumber \\
	   & - U_{v}^{\mu}(\mid \vec{p}_{1}\mid,x)
	   -    U_{v}^{\mu}(\mid \vec{p}_{2}\mid,x)    \label{defde6}
\end{eqnarray}
we obtain in shorthand form
\begin{equation}
\Pi^{\mu}_{1} + \Pi^{\mu}_{2}  = \Pi^{\mu}_{3}  + \Pi^{\mu}_{4} +
\Pi^{\mu}_{5}  + \Pi^{\mu}_{\bar{p}} + \Delta^{\mu}. \label{gl692}
\end{equation}
Similar relations are derived for the $\pi N$ 
production channels
where we assume the pion selfenergy to be identically to zero. 
This assumption
might appear questionable~\cite{Ymazaki1,Yamazaki2,Waas}, 
however, it has been found in a couple of studies
that the pion dynamics do not indicate strong potentials 
in the nuclear medium~\cite{Ehe,Ehehalt}.

In order to derive an expression for the differential 
$\bar{p}$-multiplicity
we assume, as in refs. \cite{Danielewicz90,Batko91,Shor90}, that the
differential elementary $\bar{p}$ production cross section is
proportional to the phase-space available for the final state in 
BB reactions:
\begin{eqnarray}
E_{3} E_{4} E_{5}E_{\bar{p}} \frac{d^{12}
\sigma_{BB \rightarrow NNN+\bar{p} }
\left(\sqrt{s} \right)}{ d^{3} \Pi_{3} d^{3} \Pi_{4} d^{3} \Pi_{5}
d^{3} \Pi_{\bar{p}}}  =  \sigma_{BB \rightarrow NNN + \bar{p}} (\sqrt{s})
\frac{1}{16 \, R_{4} ( \sqrt{s})}  \nonumber \\
    \delta^{4}
   ( \Pi_{1}^{\mu} +\Pi_{2}^{\mu}- \Pi_{3}^{\mu}
 -\Pi_{4}^{\mu} -\Pi_{5}^{\mu}
 -\Pi_{\bar{p}}^{\mu}-\Delta^{\mu}).
\label{diff12}
\end{eqnarray}
and
\begin{eqnarray}
E_{3} E_{4} E_{\bar{p}} \frac{d^{9}
\sigma_{\pi B \rightarrow NN+\bar{p} }
\left(\sqrt{s} \right)}{ d^{3} \Pi_{3} d^{3} \Pi_{4} 
d^{3} \Pi_{\bar{p}}}  =  \sigma_{\pi B \rightarrow NN + \bar{p}} 
(\sqrt{s})
\frac{1}{8 \, R_{3} ( \sqrt{s})}  \nonumber \\
    \delta^{4}
   ( \Pi_{1}^{\mu} +\Pi_{2}^{\mu}- \Pi_{3}^{\mu}
 -\Pi_{4}^{\mu}  -\Pi_{\bar{p}}^{\mu}-\Delta^{\mu}_{\pi})
\label{diff12a}
\end{eqnarray}
in case of $\pi B$ reactions.

Here, the $\delta$-functions guarantee the energy and 
momentum conservation and
$\sqrt{s}$ is the invariant energy available for the quasi-particles in
the initial state. $R_{4}(\sqrt{s})$ ($R_3$) is the 4-body (3-body) 
phase-space integral
\cite{Byckling}; it has been included to ensure that the differential
cross sections are normalized to the total cross section.

\section{Elementary reaction cross sections}

\subsection{$\pi N \to {\bar p} X$}

Since there are  no experimental data on 
the antiproton production yields in  pion induced reactions close to the 
reaction threshold~\cite{LB},  we have to address to microscopic models
in order to obtain closer constraints on this quantity.
Here we calculate the cross section of the reaction
$\pi N \to \bar p  N N$ within the One-Boson-Exchange model (OBE) using 
the OBE parameters proposed in~\cite{Holinde,Machleidt}.
Though the OBE results depend on the  cut-off parameters $\Lambda_i$ 
in the formfactors - which leads to an uncertainty in the
$\bar{p}$ production cross section by about a factor of 2 according to our
investigations - we here adopt the
results of the Bonn-J\"ulich model~\cite{Holinde,Machleidt} 
for a leading order
computation.
As in Refs. \cite{LRC,Sibirtsev1} we will complement 
the inclusive production
of antiprotons at high energies by calculations within 
the LUND string model~\cite{Lund} (LSM)
to include additionally a multi-hadronic
final state with more than two nucleons and an antiproton 
in the final state.

\begin{figure}[h]
\psfig{file=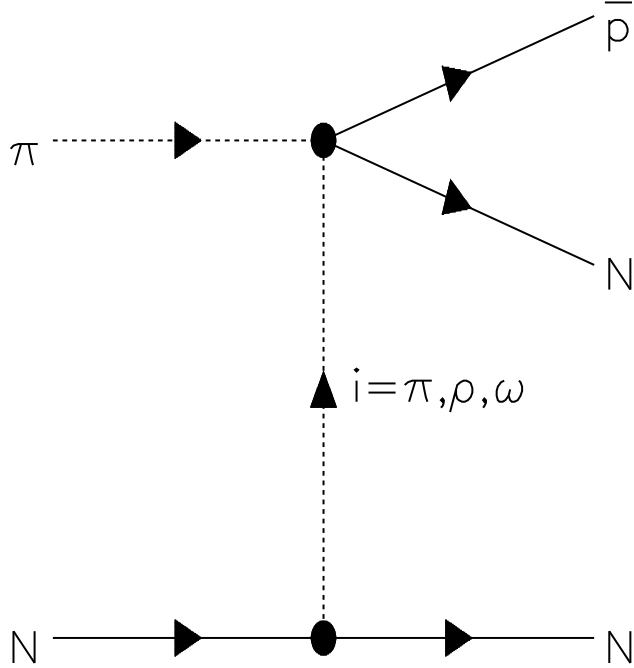,width=12cm}
\caption{\label{pibar5}Processes contributing to antiproton
production in $\pi N$ collisions.}
\end{figure}

Fig.~\ref{pibar5} illustrates  the relevant diagrams for the 
exclusive reaction $\pi N\rightarrow {\bar p} N N$ which we 
will compute in
the OBE approach. We take into account both pseudoscalar
and vector mesons exchanges and express the matrix element as
\begin{equation}
\label{1}
T_{\pi N\rightarrow {\bar p} N N}(t,s)=
\sum_i \frac{\bar u(p') \  \Gamma_i  \ u(p) \ F_i(t)}
{t -m^2_i} \ 
f_{\pi i\rightarrow \bar p N}(s_1) ,
\end{equation}
where $f_{\pi i\rightarrow \bar p N}(s_1)$ is the amplitude
for the process $\pi i \rightarrow \bar p N$ and the index $i$ stands
for the  exchanged ${\pi}, \rho $ or $\omega $-meson (cf. Fig.~1),
while $F_{i}$ stands for the corresponding formfactor. 
Here $u(p)$ and ${\bar u(p')}$ are the spinors
of the initial and final nucleons with $p$, $p'$ being their
four-momenta, respectively. 

In Eq.~(\ref{1}) $s$ denotes the square of the colliding energy
and $s_1$ is the square of the invariant mass of the $\bar p N$ 
pair, while $t$ is the four momentum transfer from the initial 
to the final nucleon shown by the lower vertex of the diagram in
Fig.~\ref{pibar5}.

The vertices $\Gamma_i$ have the following form
\begin{eqnarray}
\Gamma_\pi=g_{\pi NN} \ \gamma_5 , \ \ \ \ \
\Gamma_{\omega , \mu }=g_{\omega NN} \ \gamma_\mu \ \ \nonumber \\
\Gamma_{\rho , \mu }= g_{\rho NN} \ \gamma_\mu+
\frac{f_{\rho NN}}{2m} \ i \ \sigma_{\mu \nu} \ k^\nu  .
\end{eqnarray}
where $k$ is the four-momentum of the exchanged $\rho$-meson.

The $\rho N N$ vertex can be rewritten equivalently 
using the Gordon's relation for the current~\cite{Holinde}:
\begin{equation}
\label{2}
\Gamma_{\rho , \mu }=(g_{\rho NN} + f_{\rho NN}) \ \gamma_\mu -
\frac{f_{\rho NN}}{2m} \ (p+p')_\mu
\end{equation}
where $m$ is the nucleon mass,
$g_{\pi NN}, \ g_{\omega NN}, \ g_{\rho NN}$ are pseudoscalar 
and vector coupling constants of $\pi NN$, $\omega NN$ and $\rho NN$ 
interactions, respectively, while $f_{\rho NN}$ stands for the tensor 
coupling constant. 
We use the coupling constants from~\cite{Machleidt,Laget,Weise}; for 
completeness they are given in Table~1. 

Now the cross section of the reaction $\pi N\rightarrow \bar p N N$ 
can be written as
\begin{equation}
\label{3}
\sigma = \frac{1}{8 {\pi}^2 \lambda(s,m^2,m^2_\pi)}
\int_{4m^2}^{(\sqrt{s}-m)^2} ds_1 \  
\lambda^{1/2}(s_1,m^2_i,m^2_\pi) \ 
\sigma_{\pi\,i \rightarrow \bar p N }(s_1)
\int_{t^-}^{t^+} dt \
\frac{A_i \ F_{i}^2(t)}{(t-m^2_i)^2}
\end{equation}
where $m_i$ is the mass of the exchanged meson and 
\begin{equation}
\lambda(x,y,z)=(x-(\sqrt{y}+\sqrt{z})^2)
(x-(\sqrt{y}-\sqrt{z})^2) ,
\end{equation}
while
\begin{equation}
\label{4}
t^{\pm}=2m^2-\frac{1}{2s} \left[  (s+m^2-m^2_\pi)(s+m^2-s_1) \mp
\lambda^{1/2}(s,m^2,m^2_\pi)\lambda^{1/2}(s,m^2,s_1) \right] .
\end{equation}
We use the formfactors in the form
\begin{equation}
\label{5}
F_{i}(t)=\frac{\Lambda^2_i-m^2_i}{\Lambda^2_i-t}
\end{equation}
with cut-off parameters ${\Lambda}_i$  from~\cite{Machleidt,Weise}
(cf. Table~1).

\begin{table}
\label{tab1}
\caption{The cut-off parameters and coupling constants used in our
calculations.}
\begin{center}
\begin{tabular}{|c|c|c|c|}
\hline
$i$  & $\Lambda _{i}$[GeV] & $g^2_{iNN}/4\pi$ & $f_{iNN}/
g_{i NN}$ \\
\hline
 $\pi $ & 0.7 & 14.4 &  - \\
\hline
$\rho $ & 1.4 & 0.84 & 6.1 \\
\hline
$ \omega $ & 1.5 & 20 & - \\
\hline
\end{tabular}
\end{center}
\end{table}

Note that $\Lambda_\pi=0.7$~GeV is taken from~\cite{Weise} and
corresponds to the OBE approach of our present paper, while 
$\Lambda_\pi=1.3$~GeV is relevant for two-pion exchange 
corrections as considered in Ref.~\cite{Machleidt}.

In Eq.~(\ref{3}) the functions $A_i$  have the following forms:
\begin{equation}
A_\pi=g^2_{\pi NN}\mid t\mid , \ \ \ \ \  A_\omega=
 g^2_{\omega NN}\mid 2t+4m^2\mid ,
\end{equation}
\begin{equation}
A_\rho=\mid g^2_{\rho NN} \ (2t+4m^2)+f^2_{\rho NN} \
(3t+5m^2+\frac{t^2}{4m^2})+ 8t \  g_{\rho NN} \ f_{\rho NN} \mid
\end{equation}

\begin{figure}[h]
\psfig{file=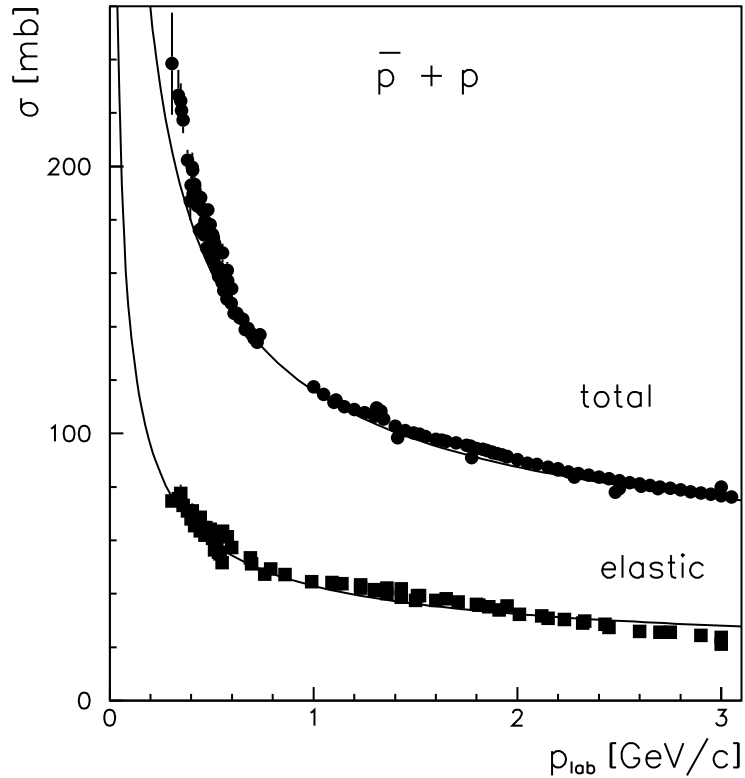,width=12cm}
\caption[]{\label{pibar4}Total and elastic cross sections for
$\bar p p$ interactions as a function of the laboratory antiproton
momentum. The experimental data are from~\protect\cite{LB}  while the
lines show the parametrization from the Particle Data 
Group~\protect\cite{PDG}.}
\end{figure}

In order to calculate $\sigma_{\pi\,i\rightarrow \bar p N}$
we address to available experimental data for the inverse reaction 
${p \bar p \to \pi\pi, \pi\rho}$ as in Ref. \cite{LRC}
which are related by detailed balance as
\begin{equation}
\sigma_{\pi i\to \bar p p}(s_1)=
\frac{4}{2J_i+1} \ \frac{s_1-4m^2} 
{\lambda(s_1,m^2_i,m^2_\pi)} \
\sigma_{p\bar p \to \pi i}(s_1)
\end{equation}
where $J_i$ is the spin of the exchanged meson and
$\sigma_{p\bar p\rightarrow\pi i}$ is  parametrized 
as~\cite{Hamer,Vandermeulen}
\begin{eqnarray}
\label{para}
\sigma_{p\bar p \to \pi  i}(s_1) = C_0 
\frac{\lambda^{1/2}(s_1,m^2_i,m^2_\pi)}
{2\sqrt{s_1}} \nonumber \\
\times 
\exp \left\lbrace -A \left[ s_1-(m_\pi+m_i)^2\right]^{1/2} \right\rbrace
\sigma^{tot}(p_{lab}) \ ,
\end{eqnarray}
where $C_0=9$~mb(GeV/c)$^{-1}$ , $A=4$~GeV$^{-1}$ and 
$\sigma^{tot}(p_{lab})$ is the total cross section for
${\bar p p}$ interactions  shown in Fig.~\ref{pibar4}. It can be 
parametrizied by~\cite{PDG}
\begin{equation}
\label{ann}
\sigma^{tot}(p_{lab})=E+B p^n_{lab}+C ln^2(p_{lab})+D ln(p_{lab}) 
\ \ \ [mb]
\end{equation}
where $p_{lab}$ is given in GeV/c and stands for the antiproton momentum 
in the laboratory system, while the parameters 
$E=38.4$, $B=77.6$, $n=-0.64$, $C=0.26$, $D=-1.2$ are taken
from the Particle Data Group~\cite{PDG}.

\begin{figure}[h]
\psfig{file=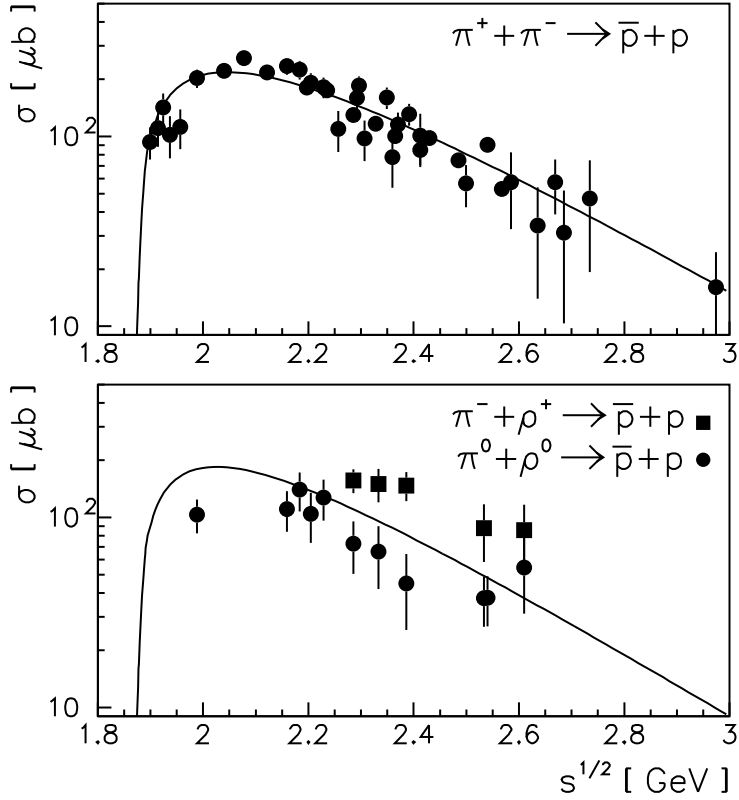,width=12cm}
\caption[]{\label{pibar1}The cross sections for the reactions
$\pi +$ meson $ \to\bar p p$. The data are extracted from the 
experimental cross sections for the inverse reaction~\protect\cite{LB};
the lines show the parametrization~(\protect\ref{para}).}
\end{figure}

Fig.~\ref{pibar1} shows the experimental data for the reaction
$\pi$+ meson $ \to \bar p p $ calculated via detailed 
balance together with
the parametrization~(\ref{para}).  In the following calculations we
assume that $\sigma_{\pi\,i\rightarrow \bar p N}$ is the same for
protons and neutrons and does not depend upon the charge of 
the exchanged meson.

\begin{figure}[h]
\psfig{file=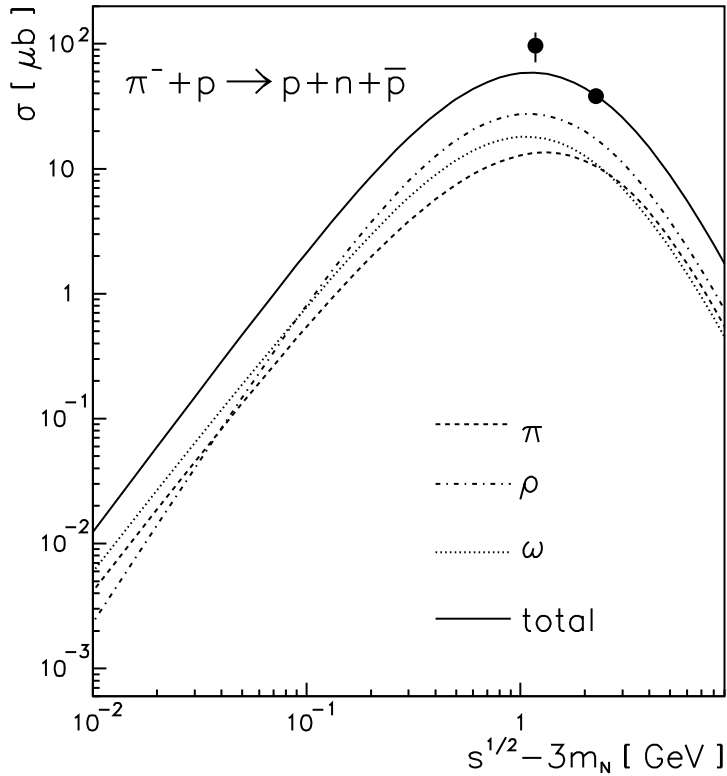,width=12cm}
\caption[]{\label{pibar2a}Cross sections for the reactions
${\pi}^-p \to p n \bar p$. 
The experimental data are from~\protect\cite{LB} while the lines indicate
our calculations within the One-Boson-Exchange model.}
\end{figure}

The cross section of the reaction ${\pi}^-p \to p n \bar p$ 
calculated with Eq.~(\ref{3}) is shown in Fig.~\ref{pibar2a} together
with the two available experimental points at higher energy. 
The solid line 
indicates the total result, while the dashed, dotted and dashed-dotted
lines show the separate contributions from $\pi$, $\omega$ and
$\rho$-meson exchanges, respectively.

\begin{figure}[h]
\psfig{file=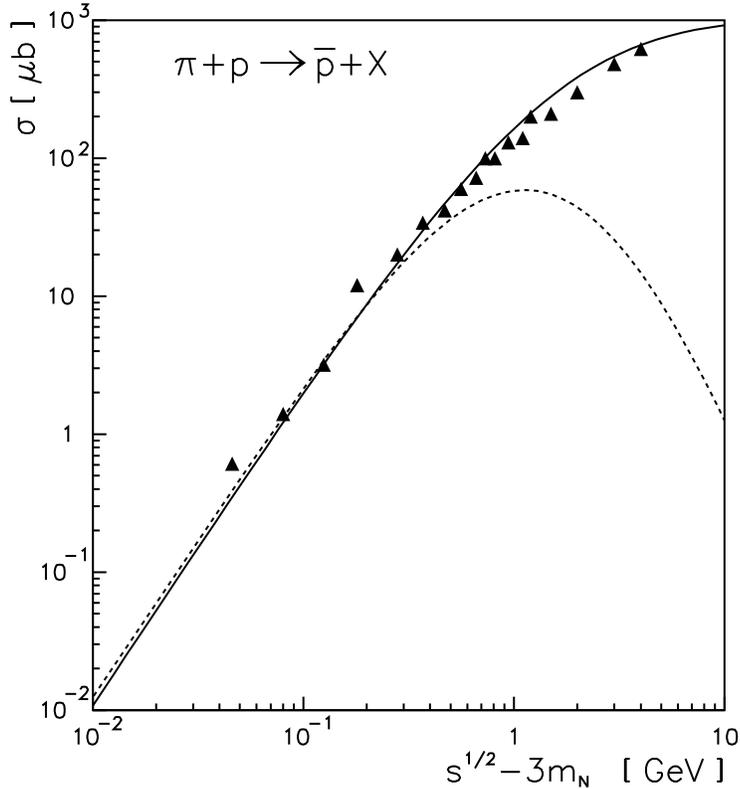,width=12cm}
\caption[]{\label{pibar2b}The cross sections for the inclusive
antiproton production from pion-nucleon collisions. The
triangles show the results from  the string model (LSM). The dashed line
is our calculation for the exclusive reaction within the 
OBE approach while the
solid line indicates the parametrization~(\protect\ref{hj}).}
\end{figure}

The inclusive cross section for the reaction 
$\pi  p\rightarrow\bar p X$ is calculated  within 
the framework of the LSM~\cite{Lund} and is shown 
in Fig.~\ref{pibar2b} by 
triangles. Here the dashed line indicates the results from the OBE
model for the exclusive channel. Fig.~\ref{pibar2b} illustrates a quite 
reasonable agreement between the results from 
two different approaches used 
in our calculations for energies close to threshold. 

For practical use in transport calculations we parametrize the antiproton
production cross section  in the form
\begin{equation}
\label{hj}
\sigma_{\pi N \to \bar p X} = a \
{\left( \frac{s}{s_0}-1 \right) }^b
{\left( \frac{s}{s_0} \right)}^c
\end{equation}
with parameters listed in Table.~2. Using the
form (\ref{hj}) we also parametrize the cross section
for antiproton production from nucleon-nucleon 
collisions as calculated within the OBE-approach in Ref.~\cite{LRC}. 
Note that the first term of Eq.~(\ref{hj})
reflects the energy dependence of the cross section near
the reaction threshold; our final results indicate
a phase-space dominance for the antiproton production cross 
sections both from
pion and nucleon induced reactions which supports  the ansatz~(23)
for the muli-differential cross section. It is clearly seen from a
comparison of Eq.~(\ref{hj}) with parameters $b$ from 
Table~2.

\begin{figure}[hbt]
\psfig{file=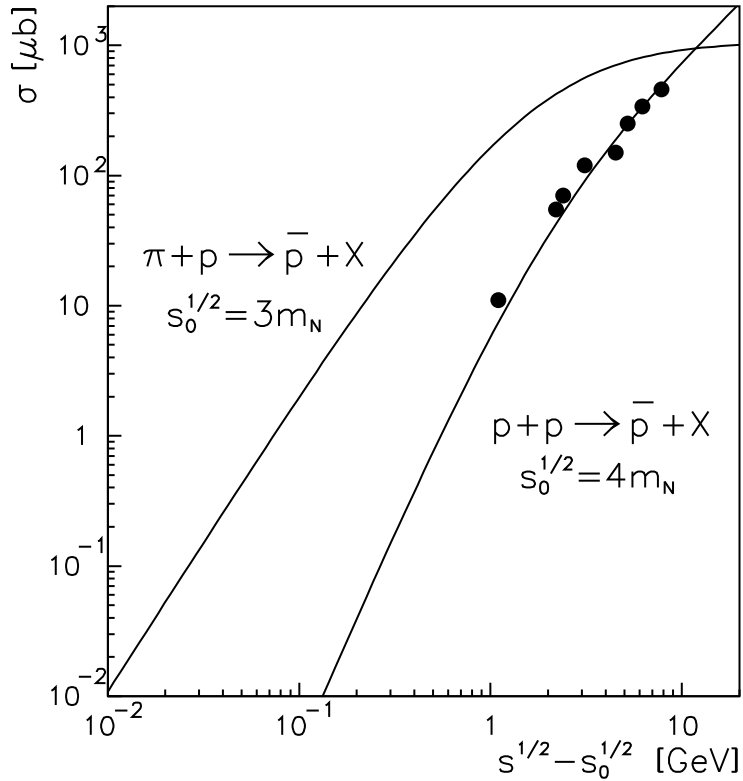,width=12cm}
\caption[]{\label{pibar3}Cross sections of inclusive 
antiproton production from $\pi p$ and $p p$ collisions. 
The circles show the experimental data for the reaction
$p p \to {\bar p} X$~\protect\cite{LB}  while the lines are the
parametrizations~(\protect\ref{hj}) of our calculations.}
\end{figure}

Fig.~\ref{pibar3} shows our parametrizations of the cross sections
for antiproton production from $\pi p$ and $p p$ collisions as
a function of the excess energy $\sqrt{s}-\sqrt{s_0}$. The solid
circles indicate the experimental data for the reaction 
$pp \to \bar p X$~\cite{LB}. We are not aware of inclusive data for
$\bar p$-production by pion induced reactions.  Note that the $\pi N$ 
channel above
threshold is much larger than the $pp$ channel since it increases with
3-body phase space compared to 4-body phase space for $pp$. This
indicates that the contribution from secondary pion induced reactions to
antiproton production in proton-nucleus and heavy-ion collisions might
become important or even of leading order.

In the following calculations we
assume that the cross sections are the same for the reactions with 
protons, neutrons and $\Delta$-resonances and for $\pi^+$, 
$\pi^-$ and $\pi^0$-mesons. 

\begin{table}
\label{tab2}
\caption{The parameters in the approximation~(\ref{hj}).}
\begin{center}
\begin{tabular}{|c|c|c|c|c|}
\hline
$Reaction $ & $s_0$ & a [mb] & b & c  \\
\hline
$\pi N \to {\bar p} X $ & $9 \ m^2$ & 1 & 2.31 & 2.3  \\
\hline
$ N N \to {\bar p} X$  & $16 \ m^2$ & 0.12 & 3.5 & 2.7 \\
\hline
\end{tabular}
\end{center}
\end{table}

\section{Comparison to experimental data}
Following the earlier studies on $\bar{p}$ 
production \cite{Teis1,Teis2}
we now attempt to extract information on the antiproton potentials
in the medium by comparing our transport calculations with the available
data, i.e. we try to fix the coupling constants 
$g_s^{\bar{p}}, g_v^{\bar{p}}$ empirically
first starting with $g_s^{\bar{p}} = g_v^{\bar{p}} = 0$.

\subsection{Proton-nucleus collisions}
Detailed experimental studies on antiproton production in
$p + A$ collisons at beam energies below the free $NN \to NNN \bar p$
reaction threshold were performed at KEK~\cite{KEK,Chiba1}. 
Antiprotons with momenta of 1. - 2.5 GeV/c were detected at an emission
angle of 5.1$^o$ in the laboratory and at  beam energies of 3.5, 4, 5 
and 12~GeV. The experiments show a very high production cross
section at the incident kinetic energy of 3.5 GeV, which is substantially 
below the free $NN$ threshold of 5.6 GeV.

Here we perform the analysis of antiproton production 
within the transport model described in Section 2.
For the propagation of
antiprotons in nuclear matter we account for both the absorption and 
elastic scattering. We assume that the elastic cross section is
37\%  of the total cross section; this simplifying assumption is shown
in Fig.~\ref{pibar4} by the solid line which
reasonably fits the experimental data.  

\begin{figure}[hbt]
\psfig{file=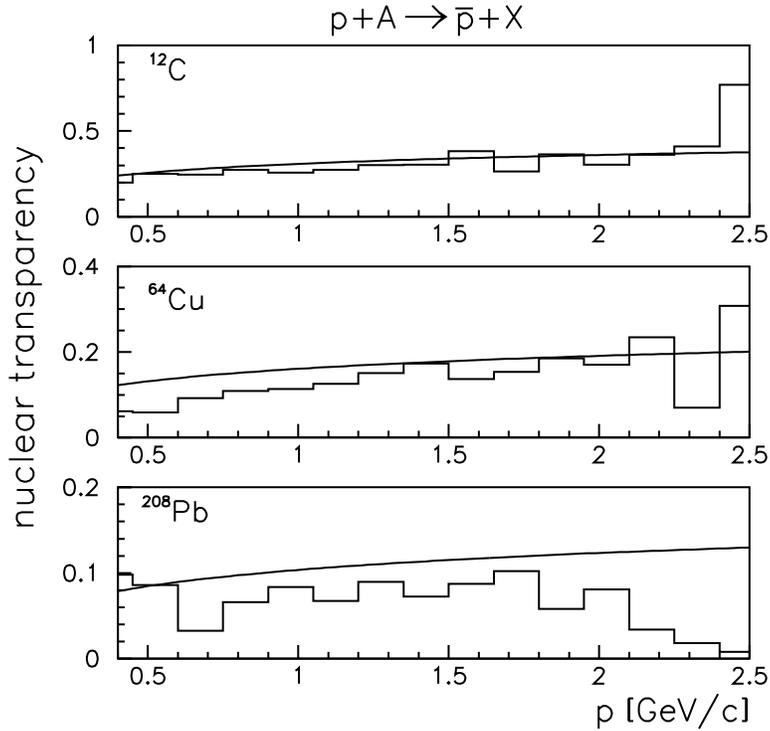,width=12cm}
\caption[]{\label{pibar7}Nuclear transparency for antiprotons produced in 
p + A reactions. The histograms show calculations within the  
transport model
while the solid lines indicate the results from the Glauber
approach in Ref.~\protect\cite{Sibirtsev1}.}
\end{figure}

The antiprotons - once produced - strongly interact with 
the surrounding nuclear
matter due to the large $\bar p N$ cross section and a large fraction
annihilates. Fig.~\ref{pibar7}
illustrates the nuclear transparency to antiprotons calculated
for different targets and defined as the ratio of antiprotons
detected asymtotically for $t \rightarrow \infty$ 
to the total number of antiprotons
actually produced in baryon-baryon and pion-baryon collisions.

The histograms in Fig.~\ref{pibar7} show our results from the transport
model while the solid lines indicate the calculations within the
Glauber formalism described in~\cite{Sibirtsev1}.
The Glauber model accounts for the interaction
of the initial protons as well as for the annihilation of the 
antiprotons. However, the rescattering in the nuclear medium 
is not included in the present version of the Glauber model.
Thus the large difference at antiproton momenta
$p \ge 2$~GeV/c between the Glauber calculations
and the results from the BUU approach illustrated in 
Fig.~\ref{pibar7} is due to rescattering effects,
which are most important for heavy nuclei.
We find that both calculations  agree reasonably well for $^{12}C$ and
$^{64}Cu$ for which experimental data are available.
Both models predict 
that for heavy nuclei  
in p + A reactions about 90\%  of the produced antiprotons annihilate in
the target nucleus.

Following the
investigations from~\cite{Teis1,Teis2} we study the influence of the
antiproton self-energy on the differential $\bar p $-production.
In the numerical calculations we adopt the 
limit $g_v^{\bar{p}} = 0$ as in 
\cite{Teis2} and vary only the scalar coupling 
$g_s^{\bar{p}}$ in Eq.~(4)
in order to obtain the best fit to the experimental data.

\begin{figure}[hbt]
\psfig{file=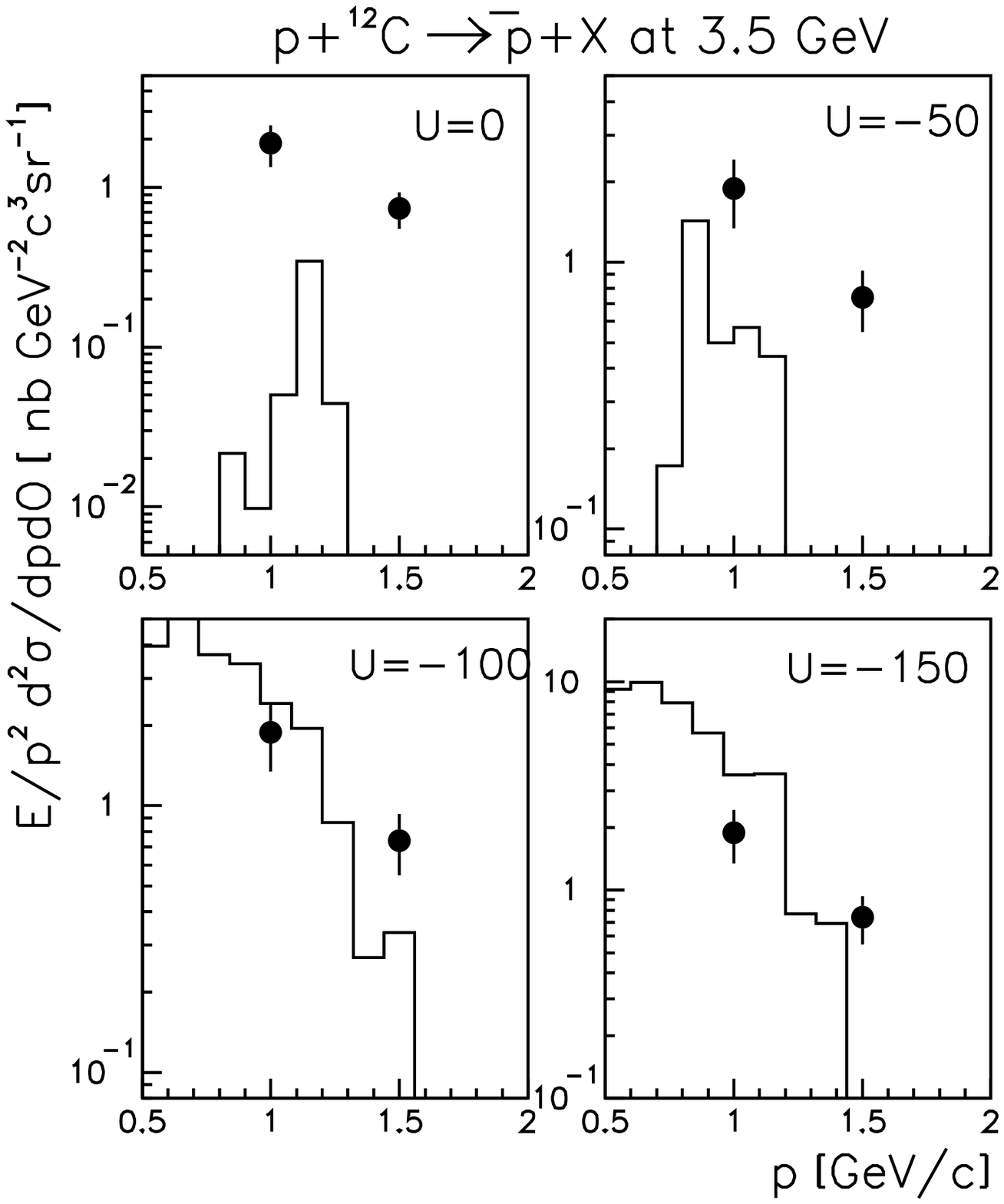,width=12cm}
\caption[]{\label{pibar6}Antiproton spectrum from $p + C$
collisions at a beam energy of 3.5~GeV. The experimental data
are from~\protect\cite{Chiba1} while the histograms indicate our 
calculations with different antiproton potentials $U$ in MeV.}
\end{figure}

\begin{figure}[hbt]
\psfig{file=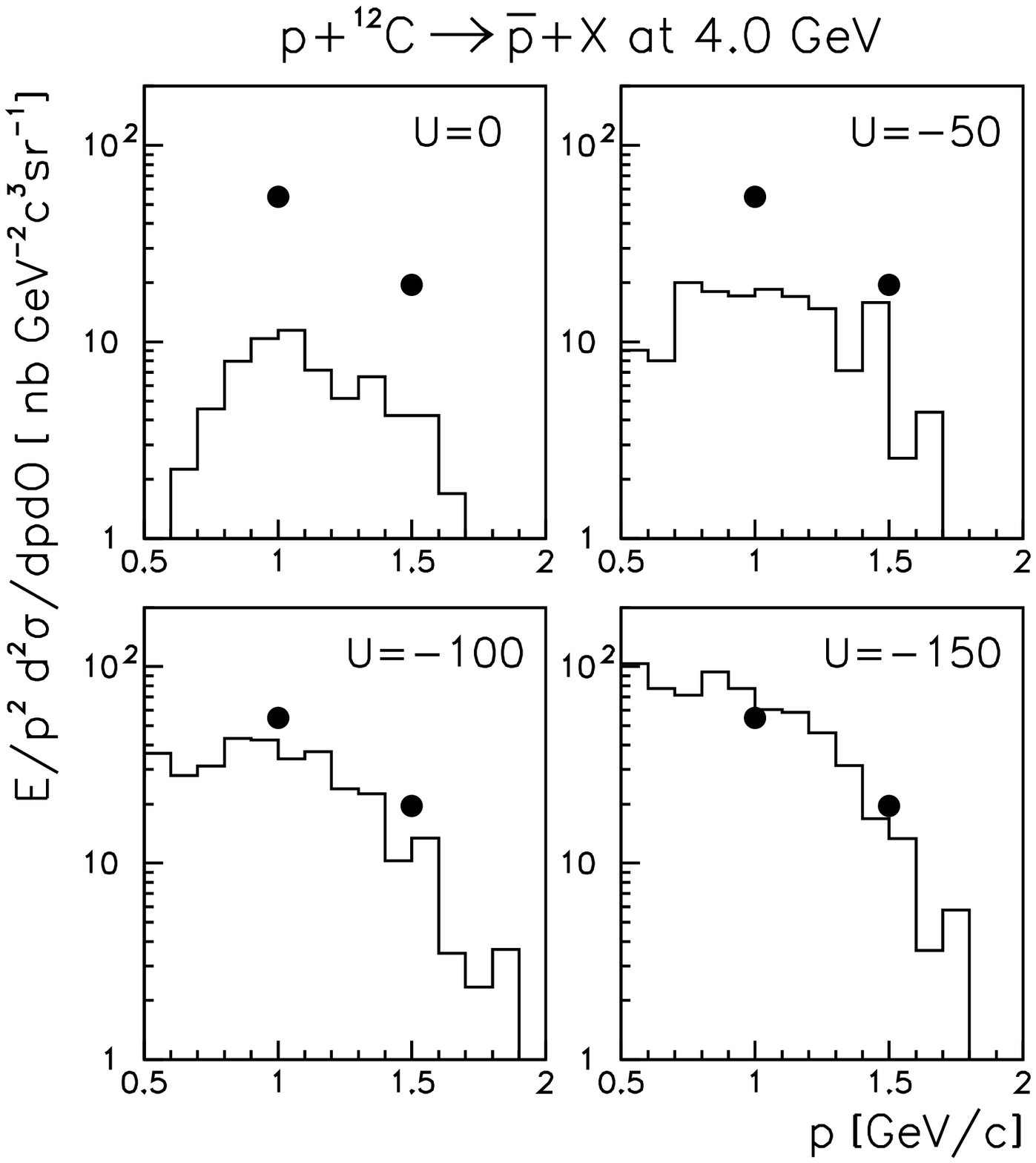,width=12cm}
\caption[]{\label{pibar6-4}Antiproton spectra from $p + C$
collisions at a beam energy of 4~GeV at $\theta_{lab} = 0^o$. 
The experimental 
data are from~\protect\cite{Chiba1} while the histograms indicate our 
calculations with different antiproton potentials $U$ in MeV at 
density $\rho_0$.}
\end{figure}

\begin{figure}[hbt]
\psfig{file=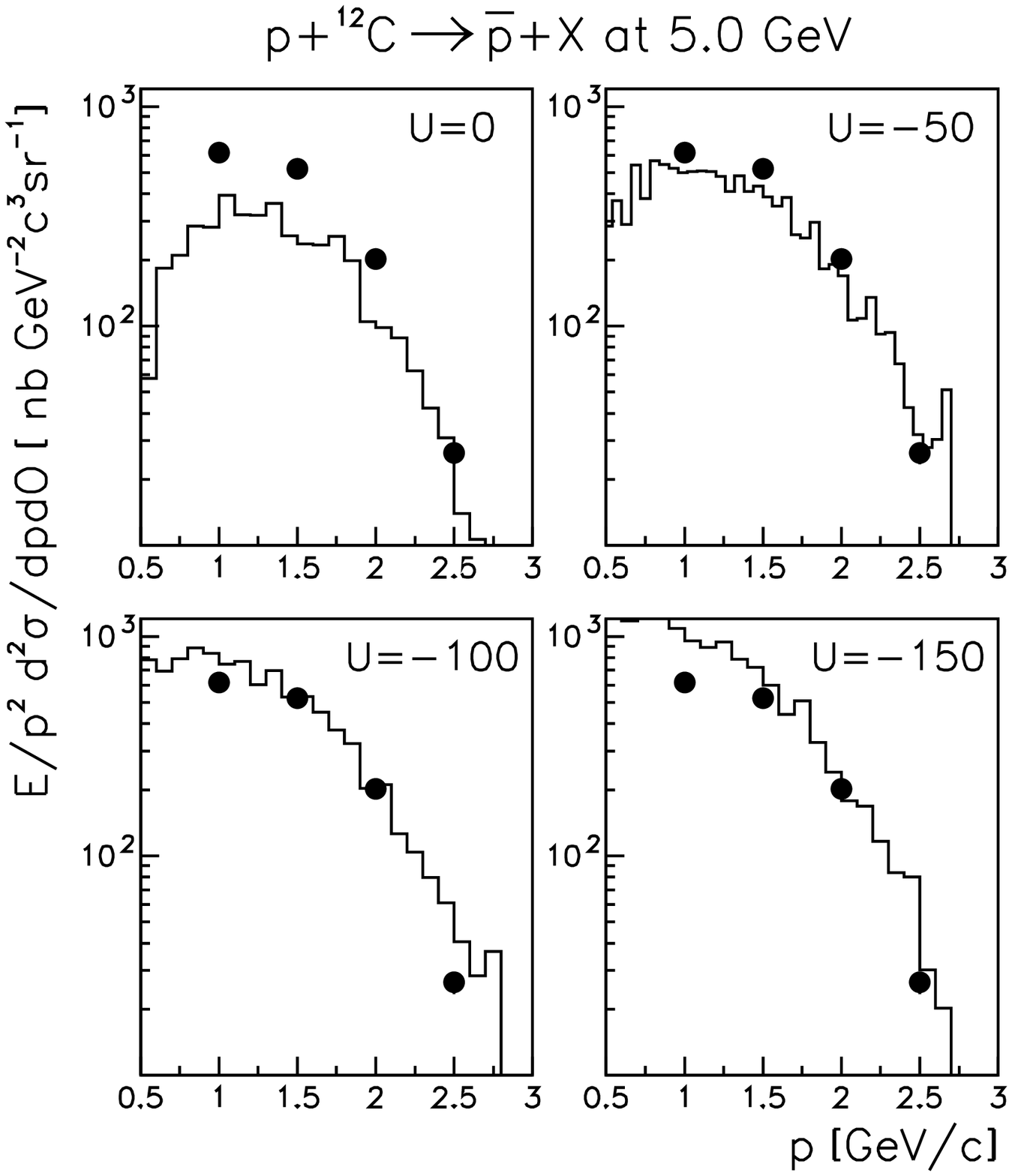,width=12cm}
\caption[]{\label{pibar6-5}Antiproton spectra from $p + C$
collisions at a beam energy of 5~GeV. The experimental data
are from~\protect\cite{Chiba1} while the histograms indicate our 
calculations with different antiproton potentials $U$ 
in MeV at $\rho_0$.}
\end{figure}

Fig.~\ref{pibar6} shows the  invariant differential cross section
for antiproton production in $p +^{12}~\!\!\!C$ collisions at  a
beam energy of 3.5~GeV and an emission angle of 5$^o$ in the 
laboratory. The 
experimental data are taken from~\cite{Chiba1} while the histograms
indicate our results calculated with different antiproton
potentials (in MeV at saturation density $\rho_0=0.16$~fm$^{-3}$).  
It is clearly seen that without in-medium
effects (U = 0) we significantly underestimate the 
production cross section.
The agreement becomes better for an attractive antiproton potential
of $-125 \pm 25$~MeV at normal nuclear matter density. 
Figs.~\ref{pibar6-4},\ref{pibar6-5}, furthermore,  show the
experimental data together with our calculations for the
reaction $p + C \to \bar p X$ at bombarding energies of 
4 and 5~GeV, respectively. Note that at higher energy the data are
no longer that sensitive to a variation of the antiproton potential.

\begin{figure}[hbt]
\psfig{file=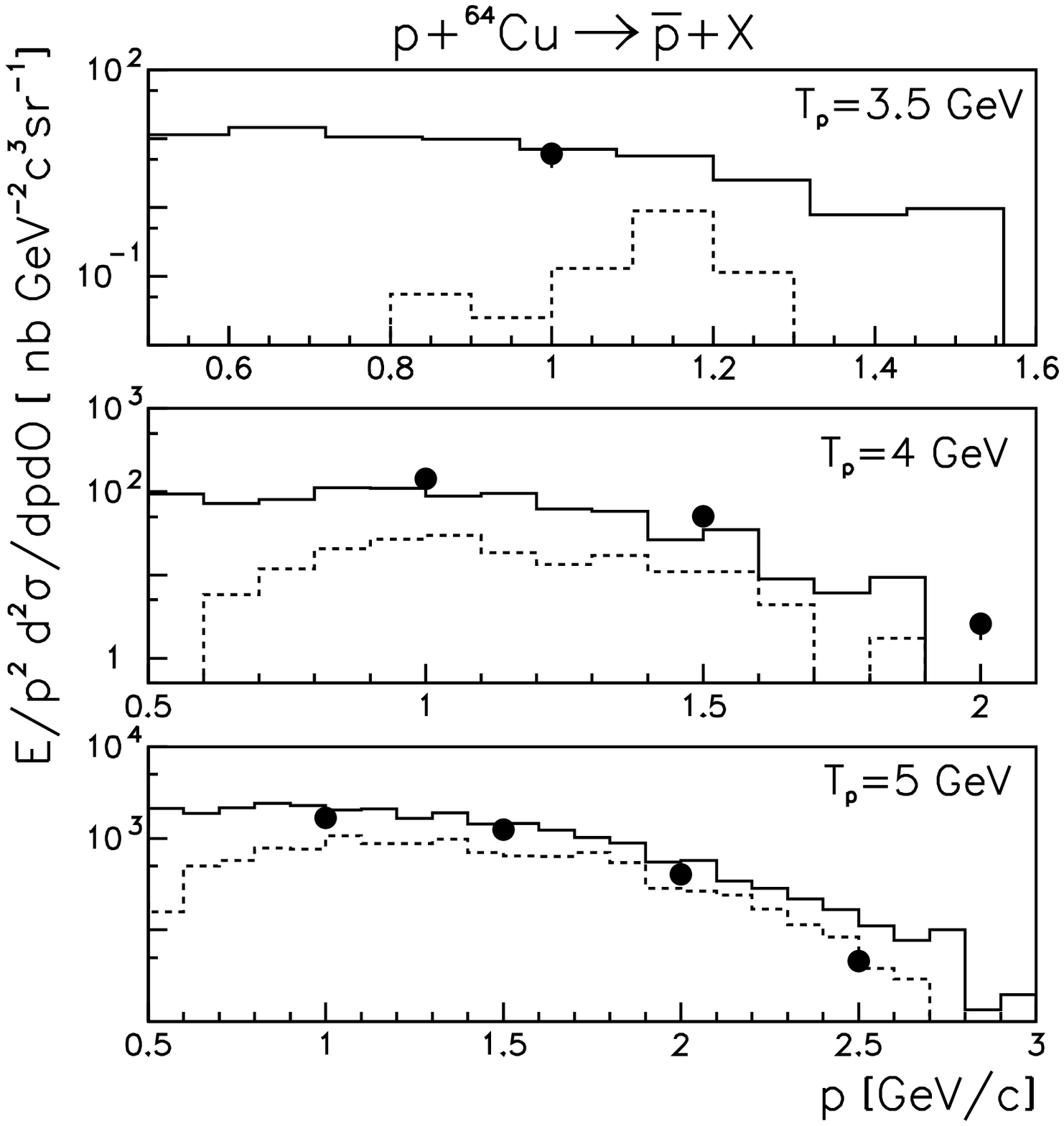,width=12cm}
\caption{\label{pibar8}Antiproton spectra from $p + Cu$
collisions at beam energies of 3.5, 4 and 5~GeV. The experimental data
are from~\protect\cite{Chiba1} while the solid histograms show our 
calculations with an antiproton potential $U=-100$~MeV at $\rho_0$; 
the dashed histograms indicate our results for $U=0$.}
\end{figure}

The antiproton spectra from $p + Cu$ collisions at
bombarding energies of 3.5, 4 and 5~GeV are shown in
Fig.~\ref{pibar8}.  The dashed histograms are the results for 
$U=0$~MeV, while the solid histograms indicate our calculations 
with an antiproton potential of $U=-100$~MeV at $\rho_0$ which 
almost perfectly 
fits the experimental data at all energies.

As anticipated before, the $\pi N$ production channel dominates
in $p+A$ reactions at least up to bombarding energies of 
6~GeV (cf. Table~3). This finding is fully in line with our
studies on $K^+$, $K^-$, $\rho $, $\omega $ and $\phi $
production for proton-nucleus reactions~\cite{Sibirtsev1}
and results from the fact $a)$ that the Fermi motion can
be exploited twice in secondary reactions and $b)$ that
the secondary $\pi N$ collisions have much larger production
cross section close to threshold (cf. Fig.~6).

\begin{table}
\label{tab4}
\caption{Cross section ${\sigma}$~[nb] for antiproton production
in $p+^{12}C$ collisions at bombarding energy $T_p$ calculated for
different ${\bar p}$ potentials $U$~[MeV].}
\begin{center}
\begin{tabular}{|c|c|c|c|c|c|c|}
\hline
$T_p$ [GeV] & Reaction & \multicolumn{5}{|c|}{ $U$ [MeV]} \\
\cline{3-7}
 & channel & 0 & -50 & -100 & -150 & -200 \\
\hline
5.0 & $\pi N \to {\bar p}$ & 362 & 563 & 824 & 1210 & 1670 \\
    & $ N N \to  {\bar p}$ & 8.4 & 20.5 & 41.7 & 78.5 & 138 \\
\hline
6.0 & $\pi N \to {\bar p}$ & 2810 & 3784 & 4895 & 6361 & 8076 \\
    & $ N N \to  {\bar p}$ & 250 & 384 & 569 & 816 & 1143 \\
\hline
\end{tabular}
\end{center}
\end{table}

In summary, we get a reasonable fit to the experimental 
data on antiproton
production from proton-nucleus collisions at subthreshold energies 
with an attractive
antiproton potential $U \simeq -125 \pm 25$~MeV at normal nuclear
matter density and at antiproton momenta from 1 to 2.5 GeV/c with respect
to the nuclear matter at rest.

\subsection{Nucleus-nucleus collisions}
Experimental studies on  antiproton production from heavy-ion
collisions at energies below the free $NN$ threshold have been
performed at the BEVALAC~\cite{Shor90,Carrol,Shor1}
and at GSI Darmstadt~\cite{GSI,Schroter2}.

Whereas in proton-nucleus collisions the antiprotons have quite large
momenta relative to the nuclear matter, in heavy-ion reactions 
the antiprotons have small momenta in the nucleus-nucleus 
center of mass and are comoving with the expanding 'fireball'. 
Moreover, heavy-ion reactions probe much higher baryon
densities, which influences both the antiproton selfenergy and their
final state interactions, i.e. reabsorption and elastic rescattering.

\begin{figure}[hbt]
\psfig{file=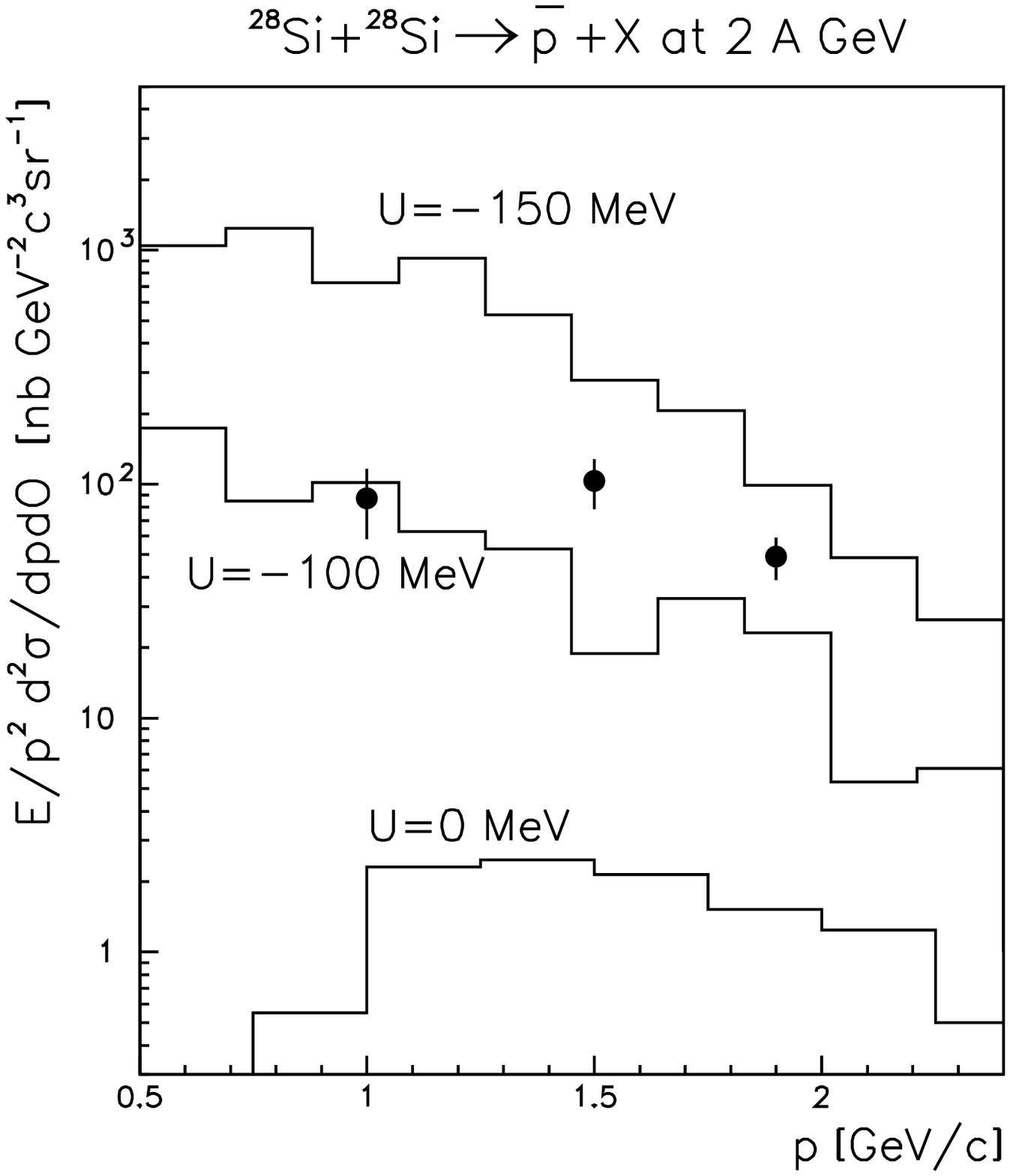,width=12cm}
\caption{\label{pibar9}Antiproton spectra from $Si + Si$
collisions at 2.0~A GeV. The experimental data
are from~\protect\cite{Shor1} while the histograms show our 
calculations with different antiproton potentials (at ${\rho}_0$).}
\end{figure}

Fig.~\ref{pibar9} shows the antiproton spectra from
$Si + Si$ collisions at a bombarding energy of 2~A GeV.
The experimental data are from~\cite{BEVALAC2,Shor1} 
while the histograms indicate
our calculations for different values of the antiproton potential $U$
(in MeV at $\rho_0$). In
Table~4 we illustrate the relative contribution to antiproton
production from $NN$, $\Delta N$ and $\pi N$ reaction channels
for a potential $U=-150$~MeV at $\rho_0$.
In order to investigate the variation of the cross section
with the impact parameter $b$ we also show in Table~4 
the differential antiproton multiplicity multiplied by the factor
$2\pi b $. It becomes clear that the dominant contribution to
$\bar p $-production also for $A+A$ collisions
stems from the secondary pion induced reactions as in case of 
$p+A$ collisions. Moreover, the dynamics of antiproton
production and propagation show no sizeable difference in the
channel decomposition as a function of the centrality of the collision.

\begin{table}
\label{tab3}
\caption{The differential antiproton multiplicity for
$Si + Si$ collisions at 2~A GeV 
multiplied by the factor $2\pi b $  for an
antiproton potential of $U=-150$~MeV at $\rho_0$. The decomposition is
performed for different impact parameter $b$ and nucleon-nucleon,
$\Delta$-nucleon and pion-nucleon reaction channels.}
\begin{center}
\begin{tabular}{|c|c|c|c|}
\hline
$b$ [fm] & \multicolumn{3}{|c|}{ $d\sigma /db $ [nb/fm]} \\
\cline{2-4}
 & $NN$ & $\Delta N$ & $\pi N$ \\
\hline
1 & 0.64 & 7.09 & 22.0 \\
2 & 1.1 & 9.5 & 28.9 \\
3 & 1.0 & 7.3 & 24.5 \\
4 & 0.67 & 3.3 & 4.7 \\
5 & 0.027 & 0.63 &  1.7 \\ 
\hline
\end{tabular}
\end{center}
\end{table}

\begin{figure}[hbt]
\psfig{file=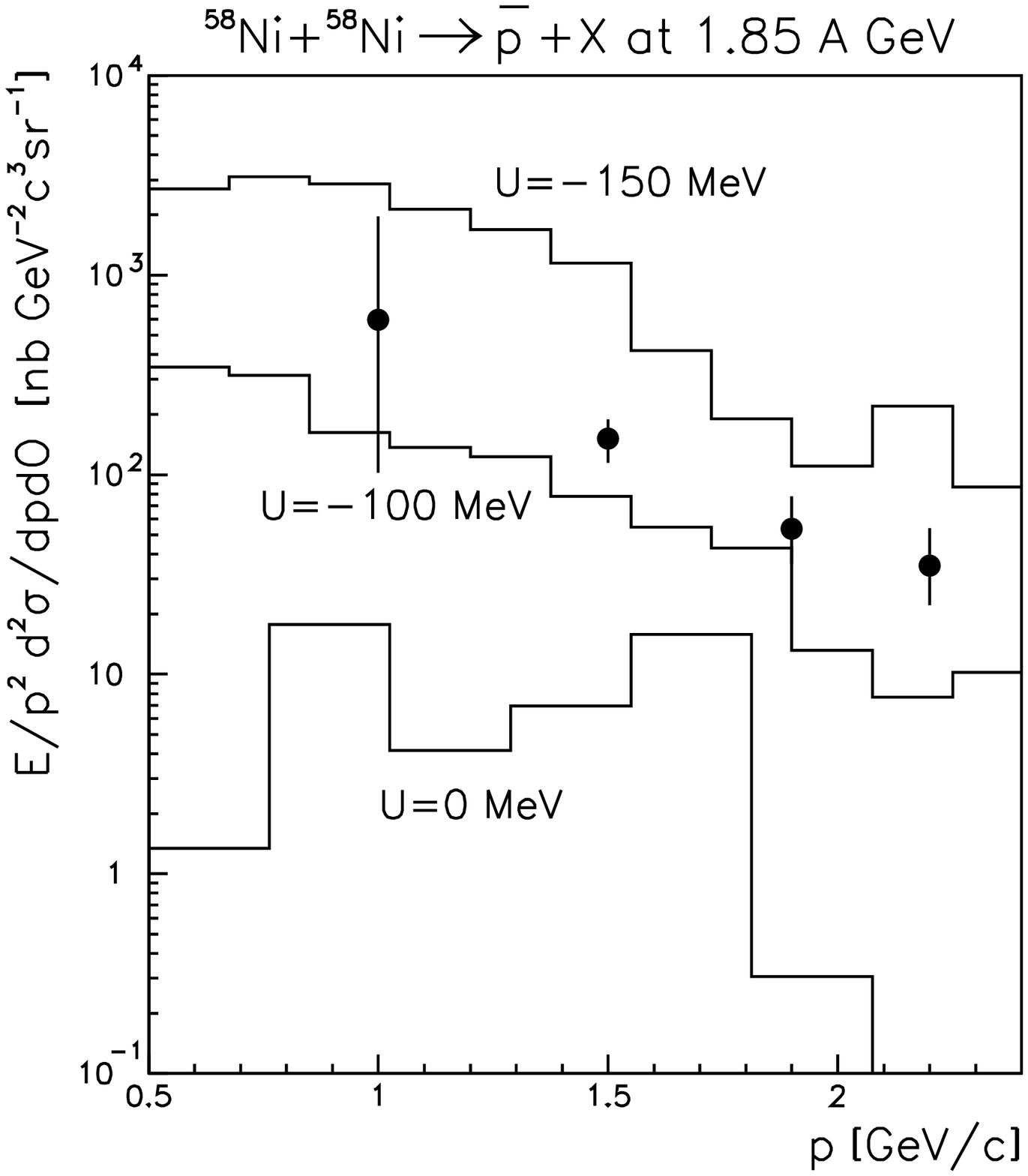,width=12cm}
\caption{\label{pibar10}Antiproton spectra from $Ni + Ni$
collisions at 1.85 A~GeV at 
${\theta}_{lab}=0^o$. The experimental data
are from~\protect\cite{Schroter2,Gillitzer} while the histograms show our 
calculations with different antiproton potentials $U$ at density
$\rho_0=0.16$~fm$^{-3}$.}
\end{figure}

\begin{figure}[hbt]
\psfig{file=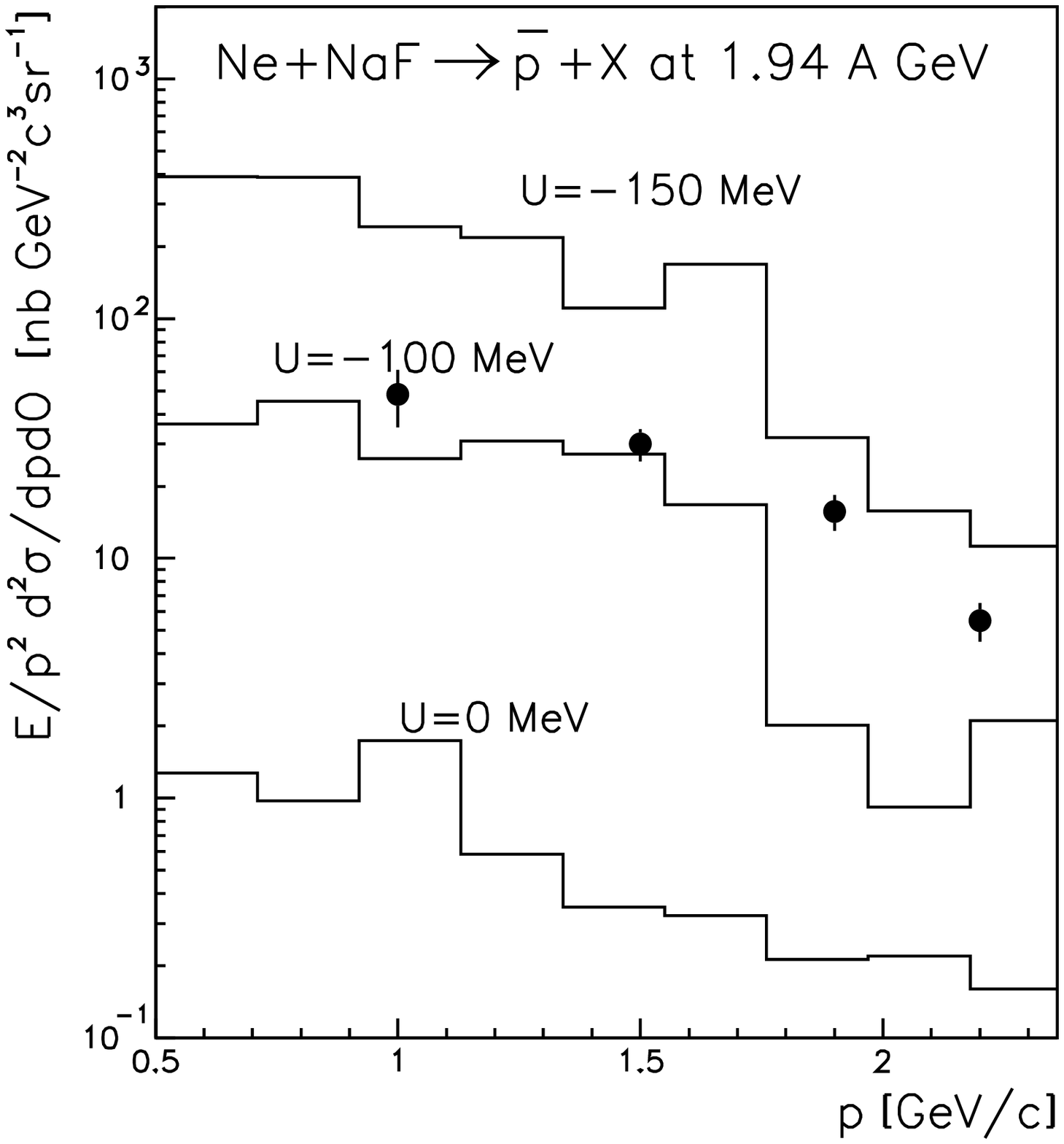,width=12cm}
\caption{\label{pibar12}Antiproton spectra from $Ne + NaF$
collisions at 1.94 A~GeV. The circles show the preliminary 
results from~\protect\cite{conference} while the 
histograms indicate our calculations for different antiproton
potentials (at ${\rho}_0$).}
\end{figure}

Fig.~\ref{pibar10} shows the antiproton spectra from $Ni + Ni$ 
collisions at 1.85~A GeV. The experimental data are taken 
from~\cite{Schroter2,Gillitzer} and can be reasonably reproduced by our 
calculations with $U \simeq -125 \pm 25$~MeV at 
$\rho_0$ whereas the data are
underestimated by about 2 orders of magnitude if no 
$\bar{p}$ potential is included.
Similar statements also hold for the antiproton spectra from 
$Ne + NaF$ collisions at 1.94~A GeV 
as shown in Fig.~\ref{pibar12} in comparison to the  preliminary 
experimental data  from~\cite{conference}.

\subsection{Dispersion approach for the antiproton potential}

The antiproton potential extracted above empirically from 
experimental data
on $\bar p$-production from $p + A$ and $A + A$ reactions now can be 
compared with the previous estimates from~\cite{Teis1,Teis2} as well as
with the optical potential from a dispersion relation approach. 
In the latter
approach the real part of the
Schr\"odinger equivalent potential in the low density approximation 
is calculated by means of the dispersion relation as
\begin{equation}
\label{disp}
U (E, {\rho}_B) = - \frac{ {\rho}_B}{2 \pi } \ 
P \int_m^{\infty} 
\frac{ dE' \ \sqrt{ E'^2 - m^2 } } { E' \ (E-E')}
{\sigma}^{ann}(E')
\end{equation} 
where $P$ stands for the principal value of the integral,
${\sigma}^{ann}$ is the  annihilation cross section
while $\rho_B$ is the baryon density.

\begin{figure}[hbt]
\psfig{file=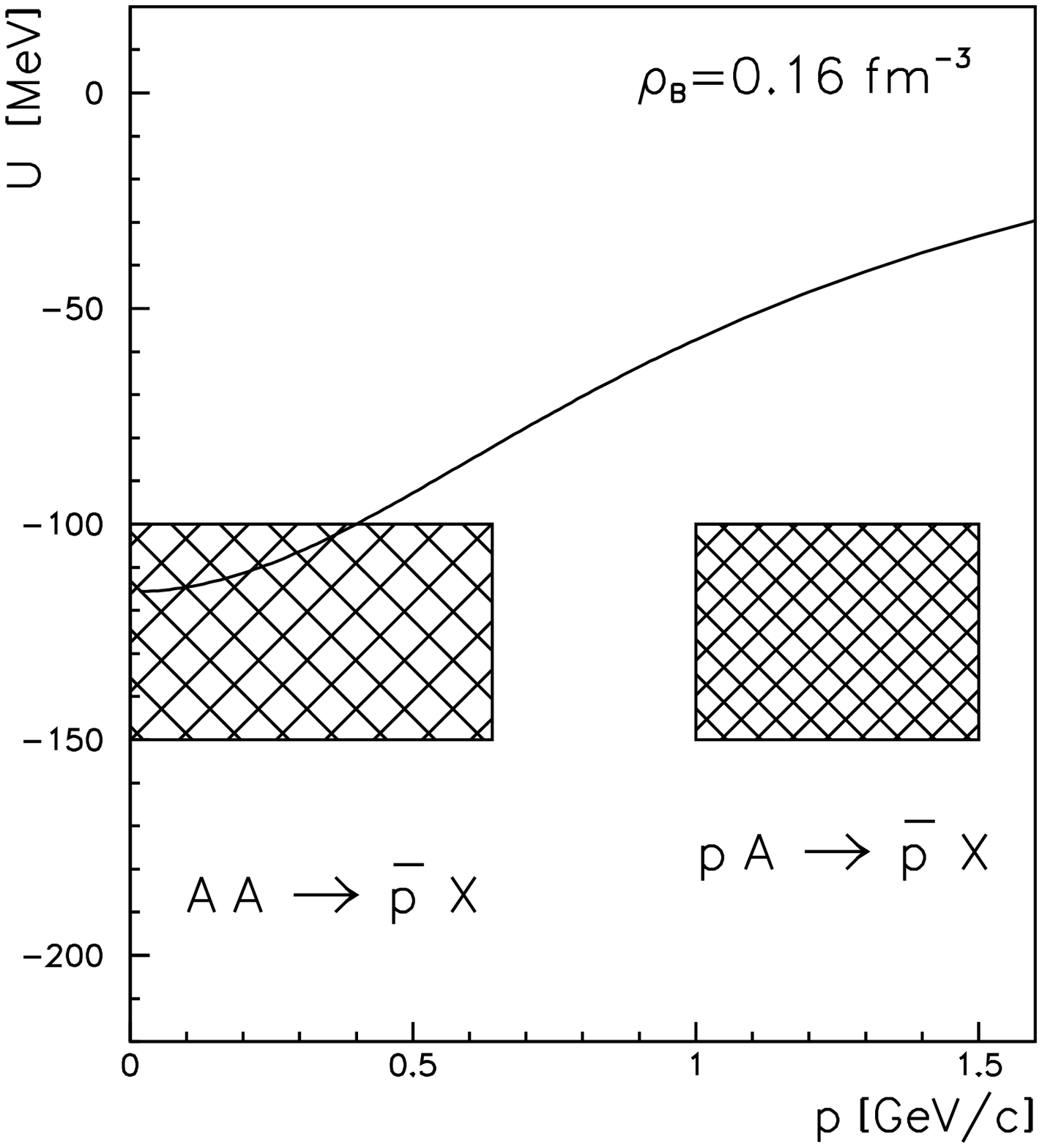,width=12cm}
\caption{\label{pibar11}Momentum dependence of the antiproton 
potential at normal nuclear density $\rho_0$. The solid line shows the
calculation with the dispersion relation (\protect\ref{disp}) 
and the annihilation
cross section from~\protect\cite{Koch89}. The rectangles are from our 
transport analysis of p + A and A + A reactions, respectively.}
\end{figure}

The solid line in Fig.~\ref{pibar11} shows the antiproton potential 
calculated by~(\ref{disp}) using the parametrization of the annihilation
cross section from~\cite{Koch89}. We note, that the integral (\ref{disp})
depends sensitively on the parametrization of the 
annihilation cross section
at high energy and that the present result using Ref. \cite{Koch89} should
be regarded cautiously.  
Fig.~\ref{pibar11}, furthermore, also  shows our results
extracted from $p + A$ and $A + A$ reactions by the rectangles, 
which are in a rough
agreement with~(\ref{disp}) at low momentum. 

\section{Summary}

In this study we have evaluated the differential cross 
section for $\bar{p}$
production for proton-nucleus and nucleus-nucleus reactions in the
subthreshold regime by considering on-shell baryon-baryon 
production channels
involving nucleons and $\Delta$'s with their in-medium quasi-particle
properties and pion-baryon collisions while treating the 
$\bar{p}$ propagation 
and annihilation explicitly.
The quasi-particle properties of the nucleons are fixed
within our approach by the nuclear saturation properties, 
the proton-nucleus
empirical potential as well as Dirac-Brueckner calculations at higher
density as in \cite{Teis1,Teis2,Ehehalt}. 
Whereas previous studies have neglected the $\pi N$ 
production channel and
used extrapolations to threshold of the elementary 
process $p + p \rightarrow
\bar{p} + X$, our approach is based on the elementary productions cross
section calculated within the OBE approach and within 
the Lund-string-model
at higher energy. These cross sections are in agreement 
with 3- and 4-body 
phase-space contrary to the extrapolation from 
Batko et al. \cite{Batko91} -
used so far in almost all studies - 
which overestimates the $pp$ channel substantially close to threshold.  

We have performed a systematic study of p-nucleus and
nucleus-nucleus collisions in a broad kinematical regime and compared our
numerical results to the data from KEK \cite{KEK} and 
GSI \cite{GSI,Schroter2}. 
We find a consistent description of all data employing an
attractive potential for the antiprotons of about -100 to 
-150 MeV at $\rho_0$ which is roughly in line with a
dispersive potential extracted from the dominant imaginary part of the
antiproton selfenergy due to annihilation. Since the dispersive model
itself is known to be valid only at small baryon density and 
the medium corrections to the annihilation cross section are practically
unknown, the agreement between our empirical analysis and the dispersion
approach is not considered to be a strong argument. 
Whereas our novel analysis roughly yields the same $\bar{p}$ 
potential at low $\bar{p}$ momenta as in Ref.~\cite{Teis2} the 
antiproton potential should be more attractive for
momenta above 1 GeV/c ($-125 \pm 25$ MeV) than assumed before.

Our analysis at subthreshold energies, on the other hand, suggests
lower antiproton potentials as anticipated from studies
at AGS energies~\cite{Koch,Spieles} where the
total antiproton production cross section is no longer that
sensitive to its value near the reaction threshold. Here
a ${\bar p}$ potential of $\simeq$-250~MeV at $\rho_0$ is
proposed in Refs.~\cite{Koch,Spieles} for an antiproton
at rest with respect to the nuclear matter, whereas an
antiproton potential of $\simeq$-160~MeV is reported for a 
${\bar p}$ momentum of 1~GeV/c~\cite{Spieles}.

Futhermore, independent from the above uncertainties it has
become clear that the $\bar{p}$ production at subthreshold 
energies is dominated by secondary pion
induced reactions as in case of kaons and antikaons \cite{Cass97,Brat97}
since for the new elementary cross sections the baryon-baryon channel
is suppressed substantially.

\vspace{1cm}
\noindent
The authors acknowledge valuable discussions with 
A. Gillitzer and J. Chiba throughout this study and
for providing us with their reanalyzed experimental 
data prior to  publication.

\end{document}